\documentclass[aps%
,preprint%
,
,pre%
,showpacs%
,superscriptaddress%
,floats%
,amssymb%
]{revtex4}

\usepackage{graphicx}
\usepackage{amsmath}
\usepackage{amssymb}
\bibstyle{apsrev.bib}

\newcommand{\comment}[1]{}

\begin{document}

\title{Tetris Model for Granular Drag}
\date{\today{}}
\author{Tomasz M. Kott}
\email{tkott@bucknell.edu}
\affiliation{Physics Department, Bucknell University, Lewisburg, Pennsylvania
17837, USA}
\author{Stefan Boettcher}
\email{sboettc@emory.edu}
\affiliation{Physics Department, Emory University, Atlanta, Georgia 30322, USA}

\begin{abstract}
Motivated by recent experiments on objects moved vertically through a
bed a glass beads, a simple model to study granular drag is
proposed. The model consists of dimers on a slanted two-dimensional
lattice through which objects are dragged very slowly to obtain full
relaxation between moves. Such an approach avoids complications due to
static friction in more realistic off-lattice models, and provides for
fast simulations at large system sizes. The upward motion of objects
of various diameters embedded in the lattice is simulated and close
resemblance with the experiments is found.
\end{abstract}
 \maketitle

\section{Introduction}
\label{introduction}
Activated dynamics of granular materials, such as sand, exhibits a
fascinating range of behaviors~\cite{Jaeger96,BTW}. For example, sand
can behave like a fluid when driven or poured from a container.  When
at rest, it adopts a solid state, such as dry sand on a beach. In the
limit of slow driving, a combination of these behaviors ensue, as
recent experiments~\cite{Koehler05,Albert99} demonstrate.

In these experiments~\cite{Koehler05}, a granular bed of glass beads
of various sizes is poured into a wide and deep dish, so that the open
surface on top is relatively smooth. Objects mounted rigidly on a
device that measures the support force are vertically plunged in or
withdrawn from the bed.  As the object is extracted, particles
displaced by the object behave as a solid, moving up en-masse.
Particles slide down intermittently, relaxing into open holes,
especially around the object itself. At short times, jamming and
avalanching, or stick-slip, motion is prevalent, while over long
periods a convective, turbulent motion occurs indicating fluid-like
behavior. Apparently similar phenomena arise when an object is plunged
slowly into a bed, although the forces on the object are significantly
larger for the relatively shallow beds considered.

The obtained data in Ref.~\cite{Koehler05} shows various interesting
scaling behaviors, regardless of the extraction or plunging of the
object. For instance, the variations of the observed average forces as
a function of depth for different object diameters $D$ can be
collapsed by a simple rescaling of the force with its buoyancy and of
the depth with $D$, independent of the grain size in the bed.

Intuitively, one would expect that plunging down through such a
granular bed would always involve larger forces than withdrawing the
same object. On the way down, the object has to work against force
chains~\cite{Liu95,Miller96,Socolar02} communicating the effect of the
rigid boundary conditions at the bottom (and sides~\cite{Vanel00}) of
the bed, while on the way up, force chains terminate at the open,
force-free boundary at the top of the bed. Hence, it comes as a
surprise that in the experiment~\cite{Koehler05} the force as a
function of the depth of the object within the bed increases with a
steeper power-law for withdrawing than for plunging. This suggests
that at large depths (much larger than attainable in those
experiments) the force of withdrawing will eventually cross-over and
exceed the force of plunging.

As the experiment (and our model here) shows, an object slowly
withdrawn from a granular bed with a two-dimensional geometry has to
lift up a bulk of interlocked material whose weight grows like
$z^\lambda$, $\lambda\approx2$, with increasing depth $z$ (measured
down from the open boundary). In contrast, it appears that a plunging
object, failing to displace any bulk of material against the rigid
boundary below, descends by {\it locally rearranging} material, for
which relevant forces merely increase with $z$ in as much as the local
pressure in the bed depends on $z$. Accordingly, in the experiment
plunging forces increase only slightly more than linear with $z$.

Unfortunately, our model does not allow for the study of plunging, for
which the consideration of frictional forces appears essential.  In
this report, we propose a simple two-dimensional model for the
withdrawal of objects from granular beds.  Inspired by the ``Tetris''
model~\cite{Caglioti97}, it ignores many characteristics of granular
materials, such as friction, to drastically simplify simulations.
Results show that the simulation model exhibits both fluid and solid
behaviors, as well as the characteristic stick-slip motion of granular
materials. Additionally, we reproduce the same data collapse found in
the withdrawal experiment, with similar scaling exponents.

\begin{figure}
\vskip 2.9truein \includegraphics{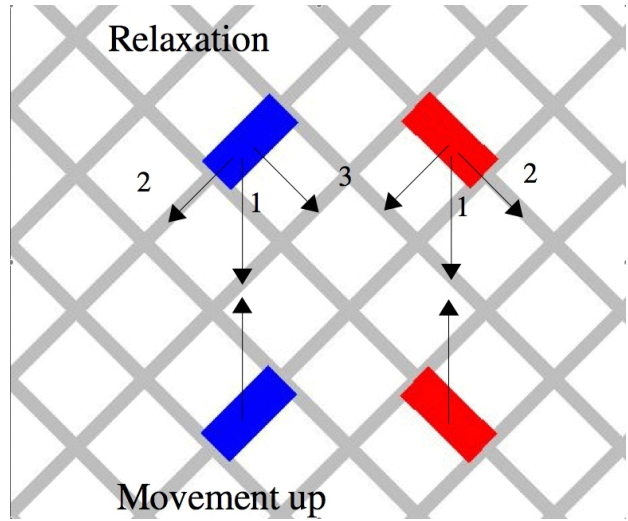}
\caption{(Color online) Depiction of the basic update steps. When
  pushed from below, dimers of either orientation can only move
  straight up one lattice unit, in turn pushing rigidly against
  everything above that occupies the targeted link, or even just one
  of the adjacent vertices. On the way down, unobstructed particles
  drop down (1) much in the same way they are moved up. Obstructed
  particles instead will slide first along their axis (2), or
  otherwise in the direction orthogonal to their axis (3). Completely
  jammed particles do not move at all. Note that a particle can only
  move when the targeted link and both adjacent vertices are
  unoccupied.}
\label{update}
\end{figure}

\section{Tetris Model for Drag}
\label{model}
In this model, dimers cover a link and the two adjacent lattice sites
and are placed on a square-lattice that is tilted by $45^o$ with
respect to the direction of gravity. These dimers, once dropped,
retain their orientation permanently in the lattice, leaning either
right or left. No two dimers can overlap, neither completely on the
link nor partially on any of the adjacent lattice sites. When pushed
or chosen for an update, dimers move according to the rules depicted
in Fig.~\ref{update}. These minimally irregularly-shaped particles
lead to complex, frustrated configurations with many uncovered sites,
merely due to the excluded volume effect.

\begin{figure}
\vskip 2.5truein
\includegraphics{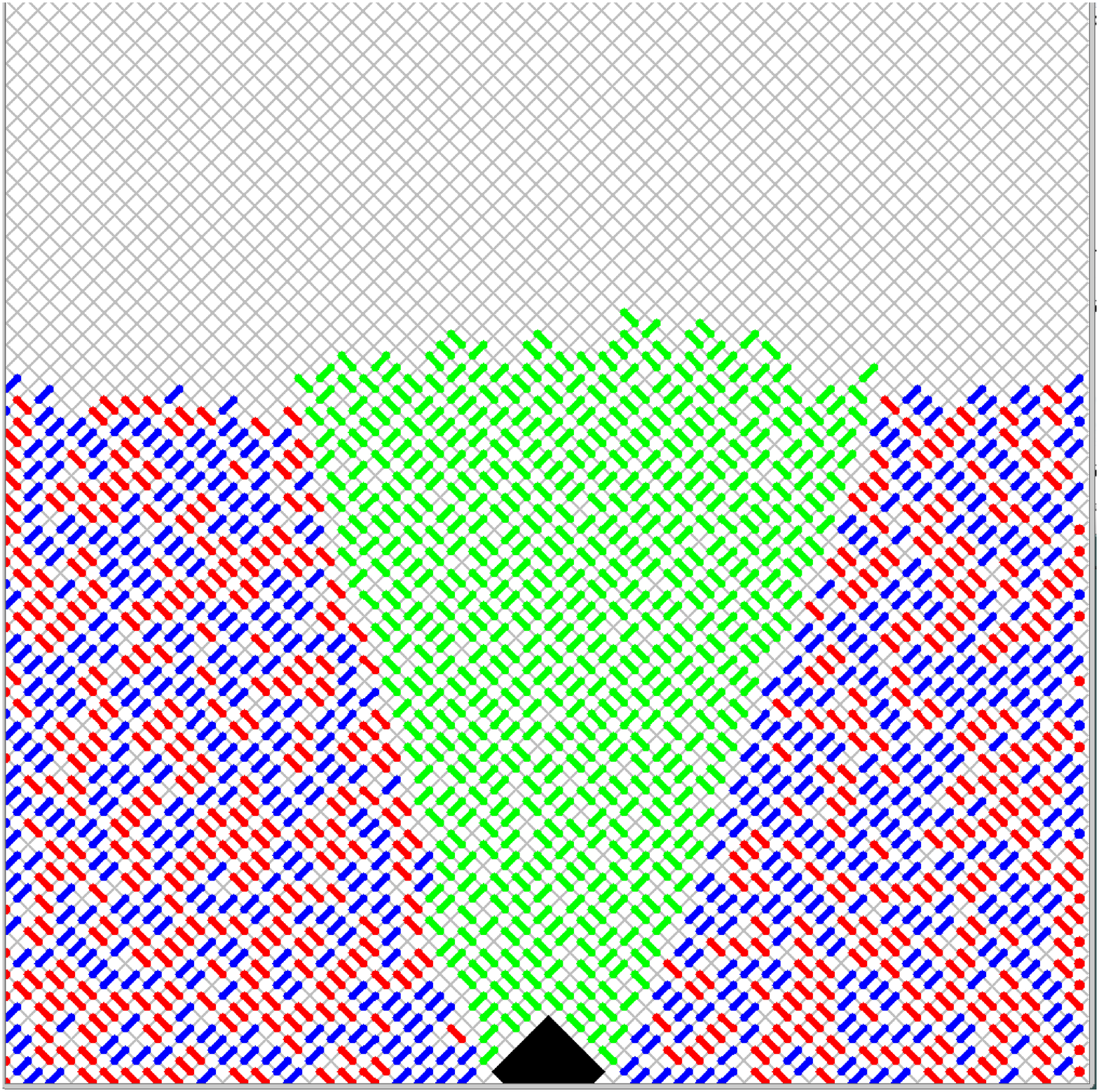}
\includegraphics{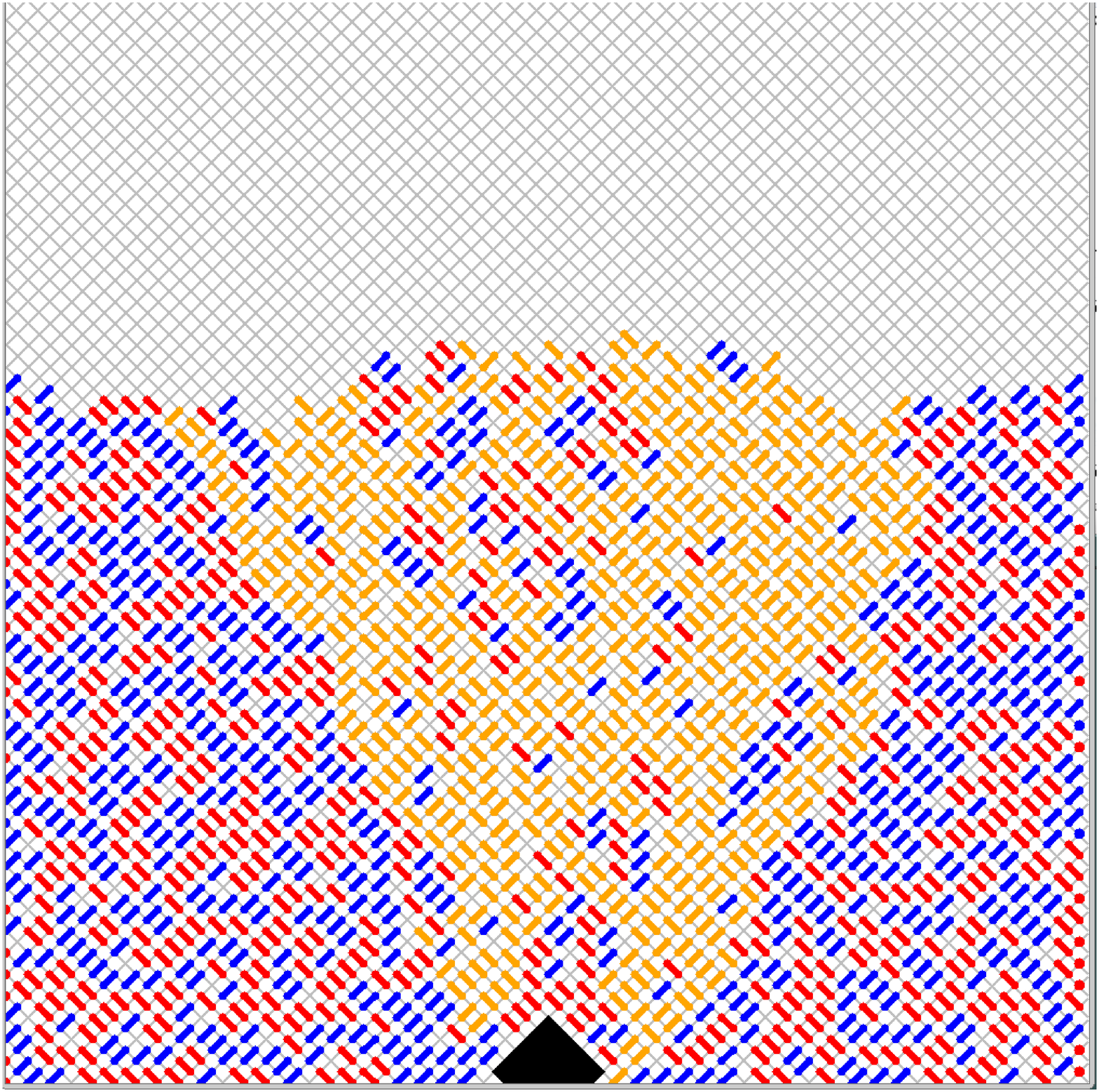}
\includegraphics{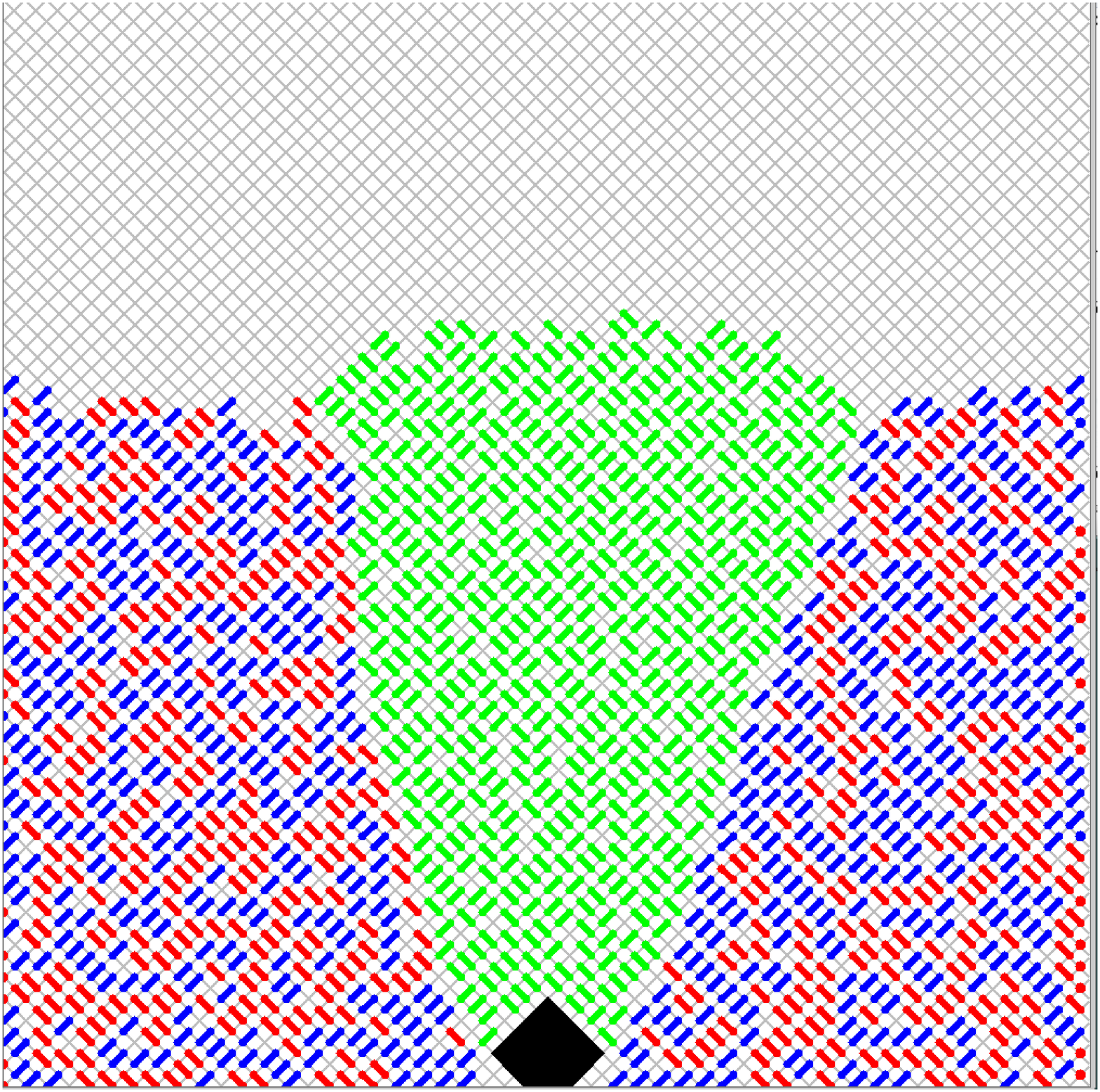}
\caption{(Color online) Visualization of the lifting of an object
(black diamond) through a granular medium of randomly arranged dimers
(red for left-leaning, blue for right-leaning) on a slanted
lattice. (Note the presence of voids in the lattice.) Green-marked
dimers on left panel are those that were forced to move when the object
displaces one lattice spacing upward. The weight of these green dimers
provides the measured force at that time-step. Note the line of voids
along the interface of the green cluster. Middle panel shows the system
after a total relaxation of all dimers that can find a void below to
slide in. This relaxation is iterated from the bottom to the top,
until no more moves are possible. All dimers that have moved during
relaxation are marked orange, showing the breadth of activity along
the interface between lifted and originally stationary
particles. Right panel shows the cluster of (green) particles being
lifted at the next time step.}
\label{lift}
\end{figure}

On the computer, we initially place dimers into the lattice, starting
from a rigid bottom, row-by-row, in random sequential order within
each row to avoid super-rough surfaces generated by purely random
deposition~\cite{Barabasi95}. At each site, dimers can initially
choose random orientations. Placements forbidden due to the excluded
volume are rejected, which leads to the formation of voids. The
system's width, which has periodic boundaries, is always chosen
sufficiently large to avoid boundary effects. Once initialized, we
insert from the bottom an object spanning a number $D$ of lattices
spacings. After each upward move (by one lattice spacing) of the
object, the granular bed of dimers is updated, following the rules in
Fig.~\ref{update}.  Fig.~\ref{lift} shows the complete cycle,
beginning with the object movement. Starting from the surface of the
object, particles are tagged if they need to move up due to a
displacement of the object or other particles below them. This process
ends only at some void or, more likely, at the open surface. The
weight of all the tagged particles (marked green in Fig.~\ref{lift})
constitute the force needed to move the object at that time. After
that force is recorded, all tagged particles are moved up in
parallel. Then, random sequentially in each row from the bottom to the
top, particles are allowed to relax by sliding into available voids
below them, according to the rules in Fig.~\ref{update}. This
relaxation process may have to be repeated, until all particles again
have a stable support. At this point, the object once again moves up
and all preceding steps are reiterated.

\section{Simulation Results}
\label{results}
\begin{figure}
\vskip 2.2truein 
\includegraphics{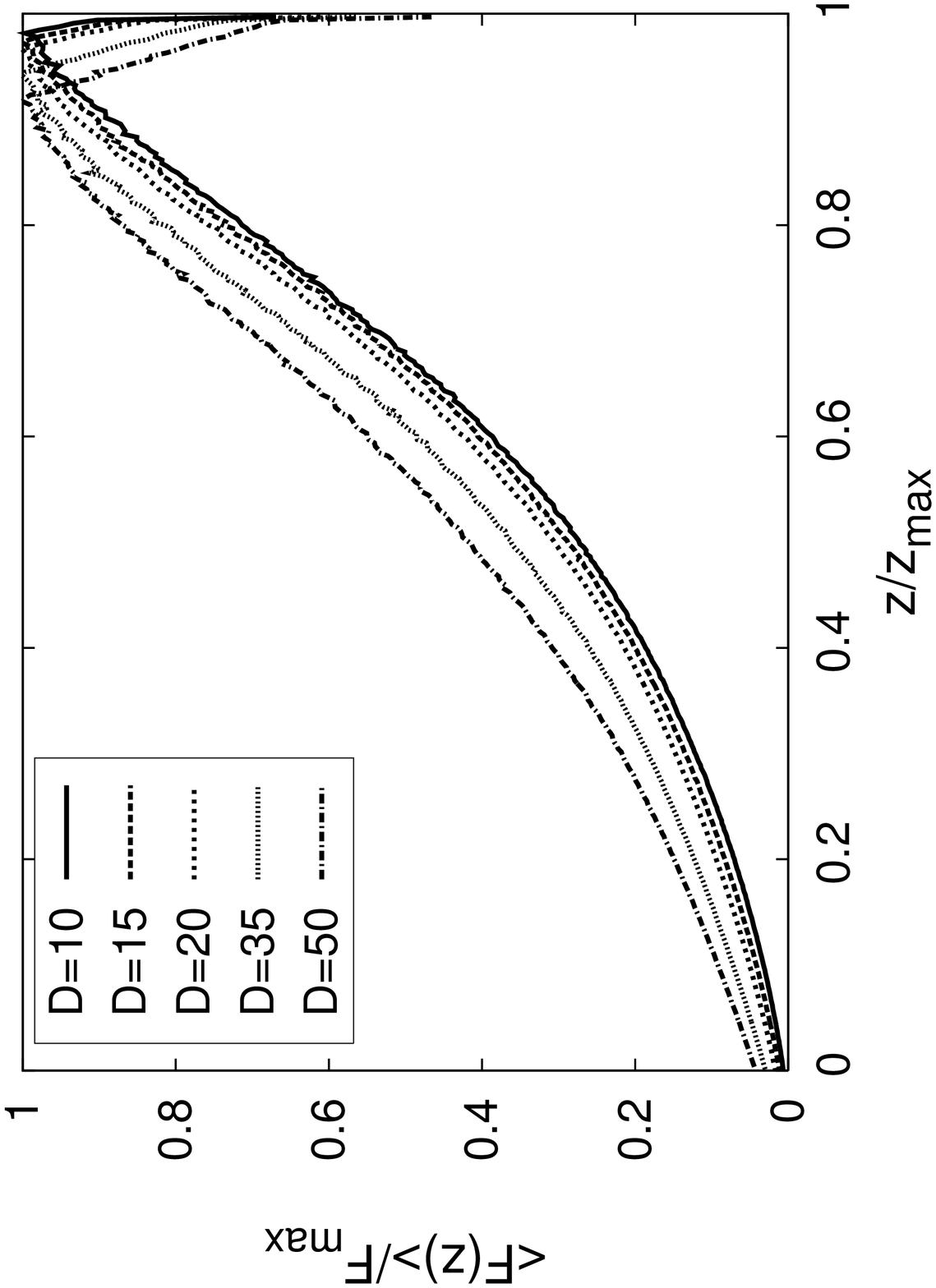}
\includegraphics{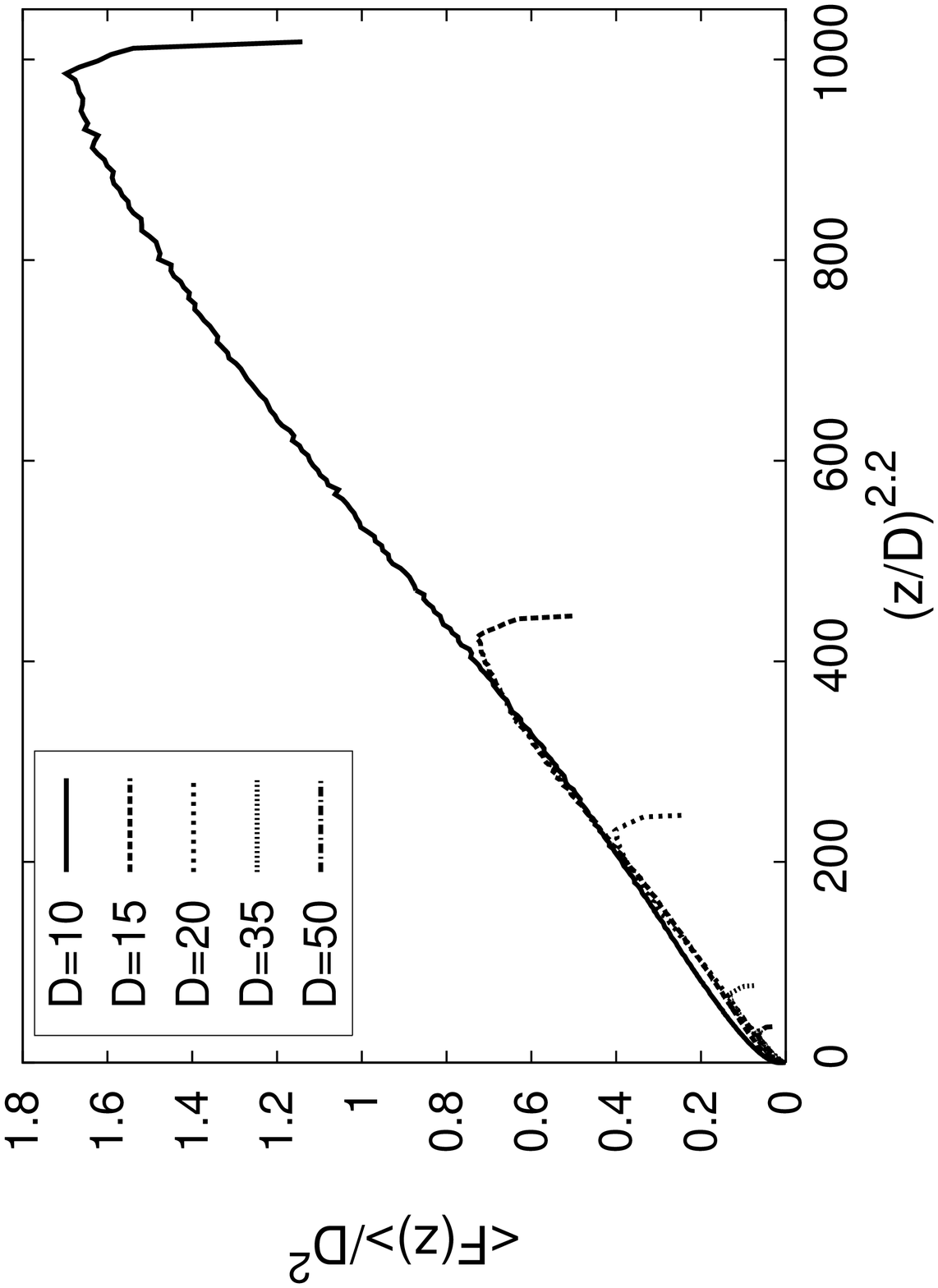}
\caption{Plot of the data for the force versus depth $z$ on objects of
  various diameter $D$ due to the lifted mass (see
  Fig.~\protect\ref{lift}). The raw data is shown on the left, and its
  scaling collapse on the right. The force, rescaled by the
  ``buoyancy'' of the object (of volume $\sim D^2$), seems to vary
  slightly stronger than quadratically with the rescaled depth, $z/D$,
  in a manner that only depends on the ratio between object size $D$
  and particle size. This was also found in the experiments, albeit
  with an apparent exponent $\lambda<2$. As Fig.~\protect\ref{lift}
  shows, the lifted cluster at each step is nearly triangular,
  i. e. of volume $\sim z^2$ (but widely fluctuating in size due to
  the disorder).}
\label{liftscal}
\end{figure}

In Fig.~\ref{liftscal} we plot the average force on the object, given
by the lifted mass, as a function of the depth $z$.  We find data
collapse of the average force vs. depth by rescaling with the object
diameter $D$ in the same way as was found for the experimental
data~\cite{Koehler05}. The force varies minutely faster than
quadratically with depth, indicating that the object lifts a wedge of
bulk material with a slightly outward-curving interface.

We let the location $z_{\cal I}(r)>0$ of the interface as a function
of the distance $r$ from the path of the object (in two dimensions)
have the functional form of a ``poweriod''~\cite{mathworld},
\begin{eqnarray}
z_{\cal I}(r)=z-Ar^{\alpha},
\label{poweroideq}
\end{eqnarray}
where $A$ and $\alpha$ are constants. In this case, the mass, the area
within poweroid, scales with the depth of the object $z^{\lambda}$,
with $\lambda={1+1/\alpha}$. In Fig.~\ref{lift}, the interface seems
to be almost linear, $\alpha\approx1$, consistent with a quadratic
relation, $\lambda\approx2$. The experiments find an exponent of
$\lambda\approx1.6$ for lifting a long horizontal bar of width $D$,
which is comparable to our two-dimensional geometry. This indicates
that the lifted wedge has a more parabolic cross-section,
$\alpha\approx1.7$. It appears that frictional forces, absent in the
model, allow steeper overhangs on both sides of the interface between
moving and stationary particles.

\begin{figure}
\vskip 2.2truein \includegraphics{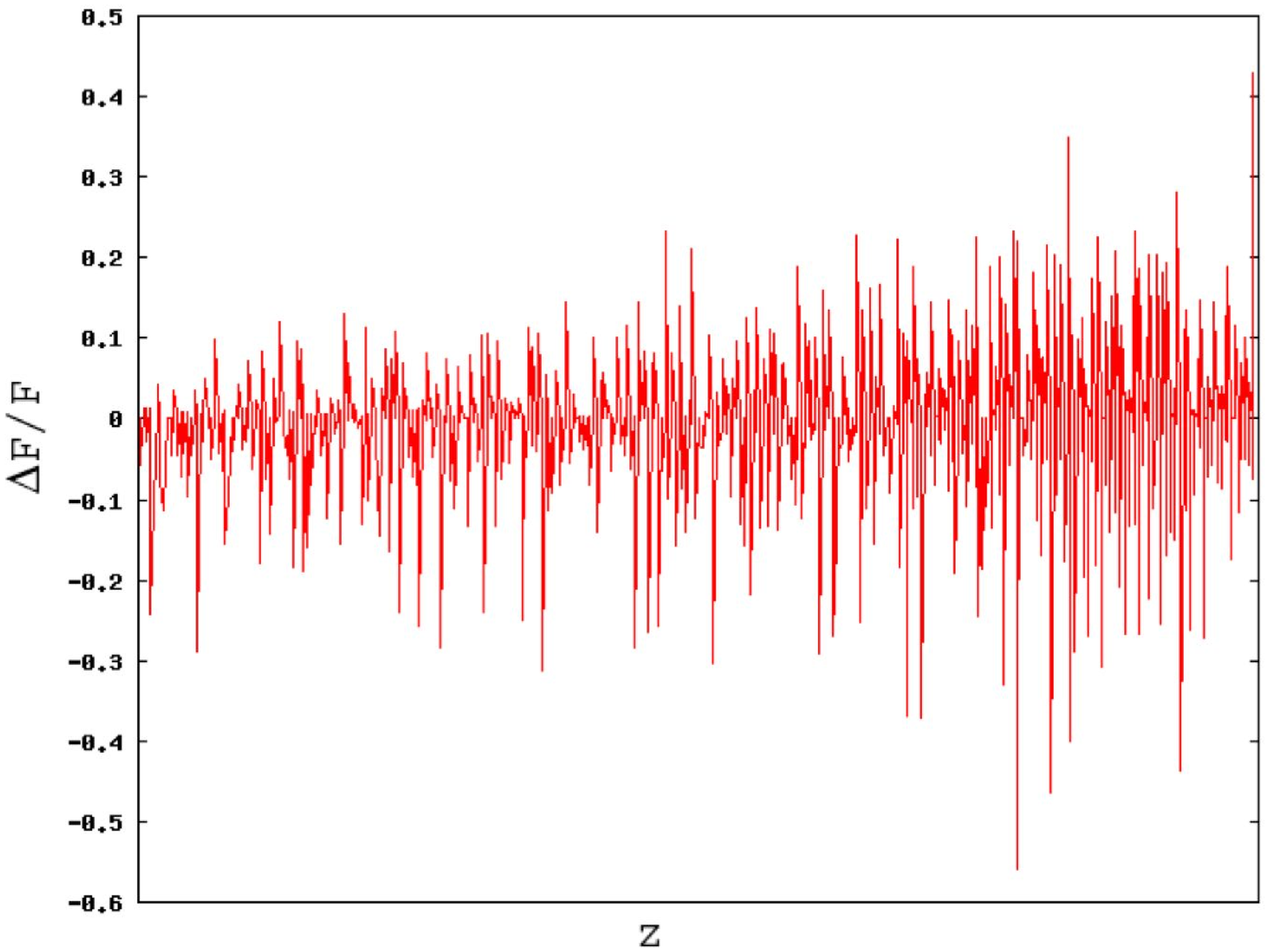}
\includegraphics{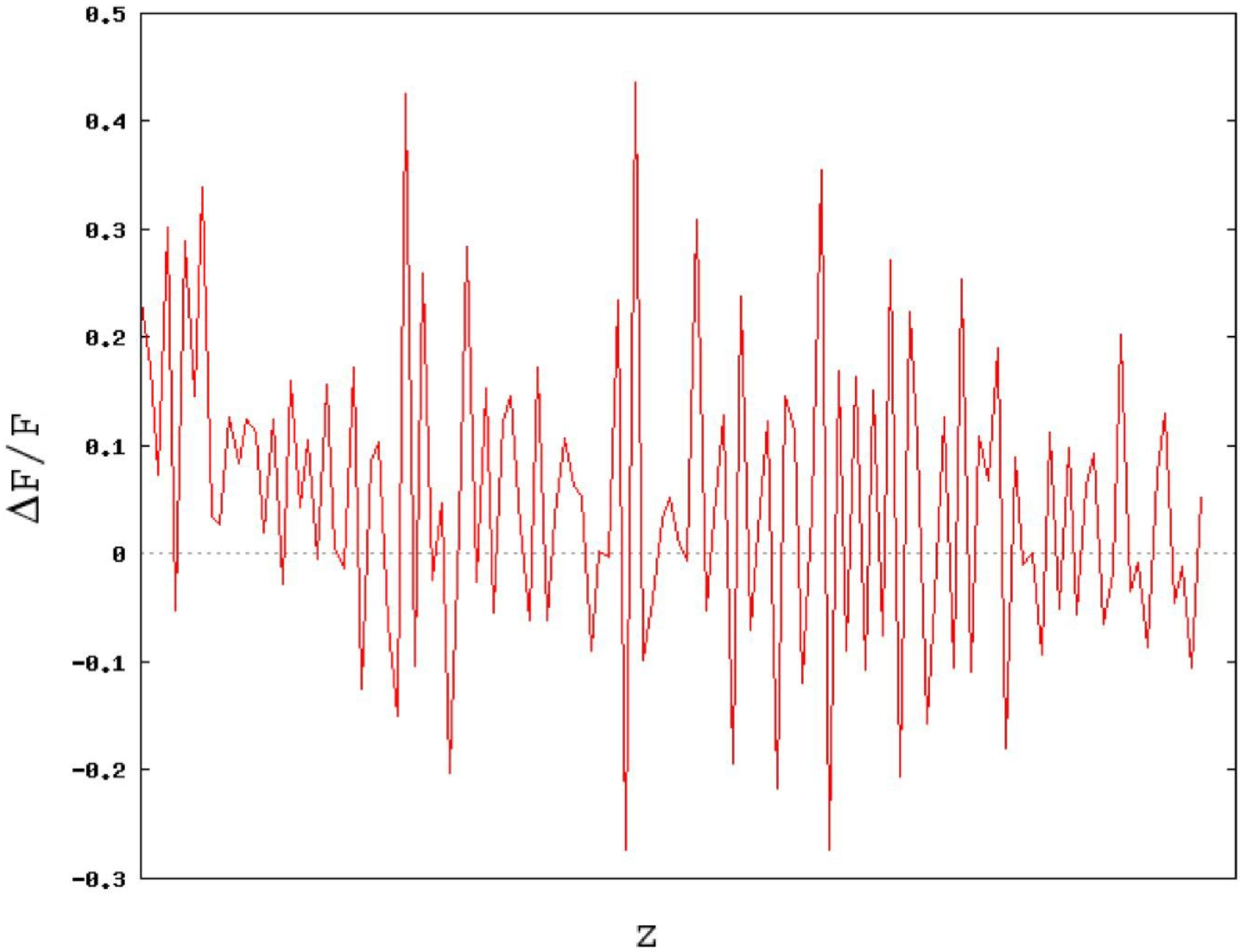}
\caption{Plot of the relative fluctuations as a function of depth $z$
  in the forces $\Delta F/F$ on the object between consecutive
  upward moves for the experimental (left) and the simulation data
  (right). Shown is the data for a generic run.  }
\label{fluctuations}
\end{figure}

Fluctuations between the forces on the object during its upward motion
seem to be entirely uncorrelated, both in the experiments as well as
in our simulations. In Fig.~\ref{fluctuations} we merely show one
generic instance of the changes in mass lifted (relative to the total
mass lifted at that point) between consecutive upward
steps~\footnote{Since in the experiment the object moves continually,
we had to discretize the data appropriately.}. Both sets of data show
the same random characteristics, and the corresponding correlation
functions are essentially trivial. This suggests that the interface
between moved and stationary particles fluctuates randomly and without
long-term memory between consecutive moves.

\begin{figure}
\vskip 1.7truein \includegraphics{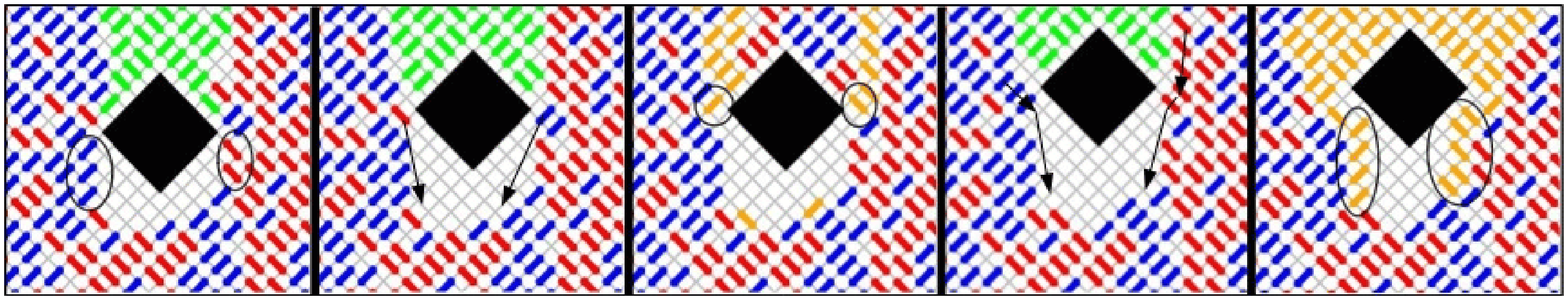}
\caption{(Color online) Sequence of motion of particles in the wake of
  a moving object. Far left, jammed particles (circled) leave a void
  behind the object. Near left, the upward motion of the object
  provides an opening for particles to spill into this void, until
  another jam ensues, shown circled in the center panel. (All moved
  particles are shaded orange.) Only in the next upward move
  (near-right panel) is this jam overcome and a significant number of
  particles pour into the void (far-right panel) to ease the weight on
  the object.}
\label{stickslip}
\end{figure}

The model also exhibits stick-slip behavior without any additional
assumptions about friction or other forces.  The irregular shape of
the particles permits the creation of voids behind the withdrawing
object by jamming the passage of other particles into the wake, as
Fig.~\ref{stickslip} shows. Thereby, the weight resting on the object
remains high (``stick'') until the jam is resolved by moving the
object further up and particles can pour in to fill the void
(``slip''). This stick-slip behavior occurs throughout a run, creating
large fluctuations in the mass lifted (see Fig.~\ref{fluctuations}).

\begin{figure}
\vskip 2.7truein \includegraphics{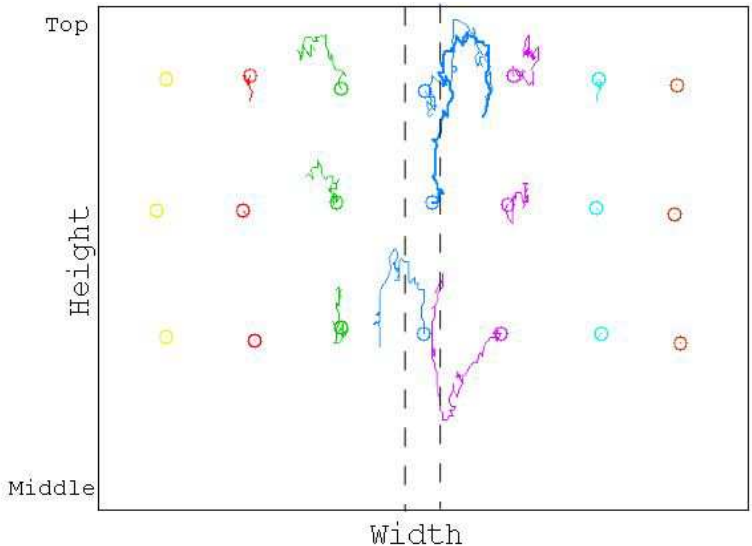}
\caption{(Color online) Plot of the motion of certain tracer particles
  during the upward motion of an object through a granular bed. Shown
  are the track, bottom to top, of the object (approximate area
  between vertical lines), the initial locations of tracer particles
  (circles) and their subsequent tracks (full lines). Even particles
  within the wedge of moved material experience only small overall
  displacements, unless they are very close to the track of the moving
  object. Those particles are first moved up and outward, then fall
  down and in toward the wake of the object (see
  Fig.~\protect\ref{stickslip}). }
\label{tracer}
\end{figure}

In addition to measuring the force on the object, we have tracked
tracer particles to study the transport of particles during the upward
motion of the object in more detail. As a result, we find fluid-like
behavior over these longer periods. As Fig.~\ref{tracer} indicates,
there exists a convective-like motion among the particles. This
convective zone does not seem to penetrate far from the track of the
(slow-moving) object.

\section{Conclusions}
We have introduced a simple model for vertical granular drag and
presented some preliminary simulation results which convey the major
aspects of granular motion, including the solid, fluid and stick-slip
behaviors. The strength of this model lies in its simplicity. Without
major assumptions about forces such as friction or electrostatics,
this simulation replicates, to good agreement, experimental data.  In
particular, the model reproduces the essential scaling collapse found
for the corresponding experiments~\cite{Koehler05}.

Future extension of this model could involve the study of fast-moving
objects, where the velocity could be tuned by prematurely stopping the
updates in the relaxation process. Another interesting application of
this model concerns motion of an object at any angle, but only above
the horizontal.  Essential for the applicability of our current model
is that any stress induced by the motion can be relieved at the open
surface.

\begin{acknowledgments}
We would like to thank Stephan Koehler for sharing his time explaining
his experiment as well as sharing his data. Tomasz Kott would like to thank
Emory University for its hospitality and the Howard Hughes Medical
Institute for supporting this research through grant \#52003727. This
work was also supported under grant \#0312510 of the Division of
Materials Research at the National Science Foundation.
\end{acknowledgments}

\end{document}